\documentclass[%
 aip,
 jmp,%
 amsmath,amssymb,
reprint,%
]{revtex4}

\usepackage{graphicx}
\usepackage{dcolumn}
\usepackage{bm}

\begin{document}



\title[Barybin-Santos]{Large-Signal and High--Frequency Analysis of Nonuniformly Doped or Shaped PN-Junction Diodes}
\thanks{One of the authors, AAB, thanks CNPq for
the support during his stay at UFPE.}

\author{Anatoly A. Barybin}
 \affiliation{Electronics Department,\\
Saint-Petersburg State Electrotechnical University,\\
197376, Saint-Petersburg, Russia.}
\author{Edval J. P. Santos}%
 \email{edval@ee.ufpe.br / e.santos@expressmail.dk.}
\affiliation{
Laboratory for Devices and Nanostructures,\\
Engineering at Nanometer Scale Group,\\
Universidade Federal de Pernambuco, Recife-PE, Brasil.
}%

\date{\today}

\begin{abstract}

An analytical theory of non\-uniformly doped or shaped PN-junction diodes submitted to
large-signals at high frequencies is presented. The resulting expressions
can be useful to evaluate the performance of semiconductor device
modeling software. The transverse averag\-ing technique is employed
to reduce the three-dimensional charge carrier transport equations into
the quasi-one-dimensional form, with all physical quantities
averaged out over the longitu\-dinally-varying cross section.
Although, it is assumed an axial symmetry, this approach gives rise 
to useful analytic expressions for the static
current--voltage characteristics, the diffusion conductance, and diffusion
capacitance as a function of the signal amplitude and the cross
section non-uniformity.
\end{abstract}


\maketitle

\section{Introduction}
\label{sec:1}

Depending on the application, different levels of device 
modeling are used to predict the behavior of electronic circuits. For 
circuits with many devices, compact models are required to reduce the 
simulation time. For circuit building blocks with a few devices, physical 
models may be used to get further insight.  Physics-based modeling requires 
more computer resources, but it has the advantage of being valid at a wider 
operating range, and offers easier to interpret parameters~\cite{SSRF2004}.
When modeling a PN-junction, different operating conditions are possible, 
such as: steady-state small-signal, steady-state large-sginal, DC, and 
transient~\cite{SSRF2004},\cite{YLY1994},\cite{RBD1995},\cite{RPFAC1993}.

The {\it transverse averaging technique\/} (TAT) allows the general 
three-dimen\-sional (3D) equations of semiconductor electronics to be converted 
into the so-called {\em quasi-one-dimensional\/} (quasi-1D) form. The 
quasi-1D equations involve all the physical scalar quantities (potential, 
charge density, etc.) and longitudinal components of the vector quantities 
(electric field, current density, etc.) in the form averaged over the 
longitudinally-varying cross section $S(z)$ and dependent only on the 
longitudinal coordinate~$z$~\cite{BS2007a}. It was first applied to 
derive analytical expressions for the depletion capacitance of the 
PN-junctions with nonuniform doping impurity profile and cross-sectional 
geometry peculiar to various real devices.   It may also be useful 
to P-$i$-N diode modeling~\cite{GSBQSB2007}. This work does not apply to 
avalanche type diodes, such as IMPATT, as no generation/ionization process 
is included in the theory at this point~\cite{SMS1981}. 

Besides, the average one-dimensional equations also include the contour
integrals along interface lines which take into account the proper boundary
conditions between different domains of the semiconductor structure.
Such equations are completely equivalent to the initial three-dimensional
equations and in this respect are accurate except that they deal with
physical quantities averaged over the cross section of a nonuniform
semiconductor structure.

This paper is devoted to the generalization of the large-signal and 
high-frequency theory of the charge carrier transport developed 
previously for uniform PN-junctions~\cite{BS2007a},\cite{BS2007b},\cite{SB2002} 
to nonuniform structures.  Application 
of the general TAT relations given in paper~\cite{BS2007a} to deriving 
the quasi-1D drift-diffusion equations for nonuniform junctions is performed 
in Sec.~\ref{sec:2}. Spectral solution of the quasi-1D diffusion equations
for nonuniform PN-junctions is set forth in Sec.~\ref{sec:3}.
Section~\ref{sec:4} deals with the derivation and analysis of the 
external circuit current, which serves as a basis for obtaining the 
static current--voltage characteristics (Sec.~\ref{sec:5}) and the 
dynamic impedance (Sec.~\ref{sec:6}) of the PN-junction diodes with 
nonuniform cross section.

\section{Derivation of Quasi-One-Dimensional Drift-Diffusion Equations
for Nonuniform PN-Junctions}
\label{sec:2}

The initial equations for derivation of the quasi-1D drift-diffusion equations
of the nonuniform PN-junctions are the three-dimensional continuity
equations and the current density expressions~\cite{SMS1981}:\\
\noindent
$\bullet $ \,\, for holes
\begin{eqnarray}
&& {\partial p\over\partial t} +\frac{1}{q}\, \nabla\cdot{\bf j}_p =
-\frac{p-p_n}{\tau_p}\,,
\nonumber\\[-2mm] &\label{eq:4.1}& \\
&&\, {\bf j}_p = qp\!\:{\bf v}_p - q\nabla (D_pp),  \quad\,\,\,
{\bf v}_p = \mu_p(E){\bf E}\!\:;
\nonumber    \qquad\quad
\end{eqnarray}
$\bullet$\,\,for electrons
\begin{eqnarray}
&& {\partial n\over\partial t}-\frac{1}{q}\, \nabla\cdot{\bf j}_n =
-\frac{n-n_p}{\tau_n}\,,
\nonumber\\[-2mm] &\label{eq:4.2}& \\
&&\, {\bf j}_n = qn{\bf v}_n + q\nabla (D_nn), \quad\,\,\,
{\bf v}_n = \mu_n(E){\bf E}\,.
\nonumber    \qquad\quad
\end{eqnarray}
with the normal components of currents that obey the following 
surface-recombination boundary
conditions~\cite{SMS1981}:
\begin{equation}
{\bf n}\cdot{\bf j}_p {\bigl |}_L = qR_{ps},
\qquad  R_{ps}=s_p(p_s-p_n),  \quad
\label{eq:4.3}
\end{equation}
and
\begin{equation}
{\bf n}\cdot{\bf j}_n {\bigl |}_L = qR_{ns},
\quad\,\,\,  R_{ns}=s_n(n_s-n_p).  \,
\label{eq:4.4}
\end{equation}
Here\, $p_n, n_p$ and $p, n$ are the equilibrium and nonequilibrium densities
of minority carriers in the $n$- and $p$-regions, $D_{p(n)},\,\mu_{p(n)}$
and $\tau_{p(n)}$ are the diffusion constant, mobility, and lifetime for
holes (electrons).

Integration of Eqs.~(\ref{eq:4.1}) and (\ref{eq:4.2}) over the cross
section area $S(z)$ by using the TAT relations~\cite{BS2007a}
produces the linear integrals along the contour $L(z)$ bounding $S(z)$. 
For holes, calculating the average and applying the 3-D extension of
Green's Theorem in (Eq.~\ref{eq:4.1}), yields 
\begin{equation}
{\partial (\bar{p}S)\over\partial t} +
\frac{1}{q}\,{\partial (\,\bar{\!j}_{pz}S)\over\partial z} +
\frac{1}{q}\!\oint\limits_{L(z)}\!\frac{{\bf n}\cdot{\bf j}_p}
{\cos\theta}\,dl = -\frac{\bar{p}-p_n}{\tau_p}\,S,
\label{eq:4.5}
\end{equation}
\begin{equation}
\frac{1}{q}\, \bar{\!j}_{pz}S =\,
\bar{p}\bar{\mu}_p\bar{E}_z S -
{\partial (\bar{D}_p\bar{p}S)\over\partial z} -\!
\oint\limits_{L(z)}\!\frac{{\bf n}\cdot{\bf e}_z}
{\cos\theta}\:D_pp\,dl.
\label{eq:4.6}
\end{equation}
Here $p_s,n_s$ are the nonequilibrium minority-carrier volume densities
taken at surface points of the structure, and $s_{p(n)}$ is the surface
recombination velocity for holes (electrons).

\begin{figure}[htb]
\includegraphics{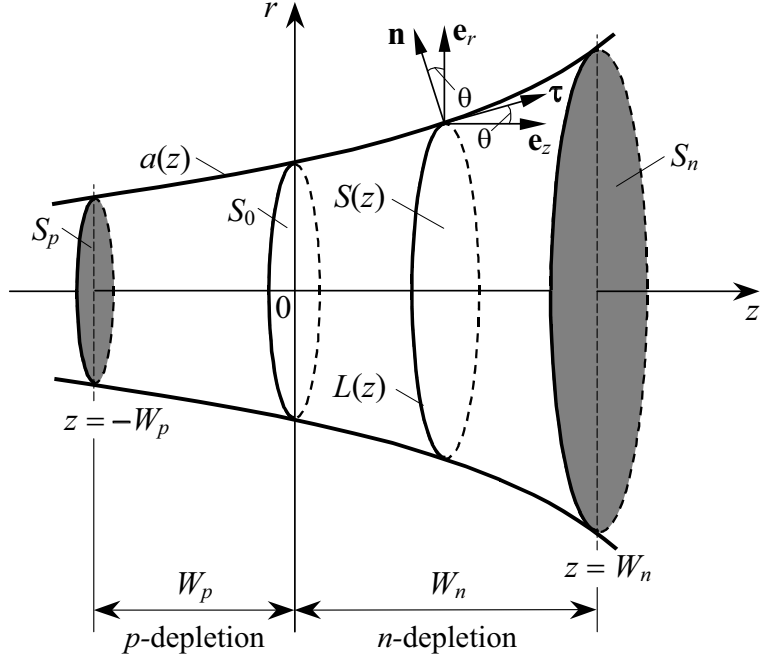}
\caption{Axially symmetric PN-junction; the depletion layer is situated between the cross sections $z= - W_p$ of area $S_p$ and  $z= W_n$ of area $S_n$.}
\label{Fig1ref5}
\end{figure}

Here the average quantities $\bar{E}_z,\,\bar{\!j}_{pz}$, and $\bar p$ are
introduced~\cite{BS2007a}, while the
average mobility and diffusion constant for holes (and similar ones for
electrons) are defined as
\[
\bar{\mu}_p=\frac{\overline{pv_p(E)E_z/E}}{\bar{p}\bar{E}_z}
\,\,\quad\mbox{and}\quad\,\,
\bar{D}_p=\frac{\overline{D_pp}}{\bar{p}}\,.
\]

For axially-symmetrical structures with $S(z)=\pi a^2(z)$, we have
\[
{{\bf n}\cdot{\bf e}_z \over{\cos\theta}} = -\tan\theta =
- {da(z)\over dz}\,,
\]
so that the contour integrals in Eqs.~(\ref{eq:4.5}) and (\ref{eq:4.6})
with the boundary condition (\ref{eq:4.3}) assume the form
\begin{equation}
\frac{1}{q}\oint\limits_{L(z)}\frac{{\bf n}
\cdot{\bf j}_p}{\cos\theta}\,dl=
2s_p\frac{\sqrt{1+ (da/dz)^2}}{a(z)}\,(p_s-p_n)S ,
\label{eq:4.7}
\end{equation}
\begin{equation}
\oint\limits_{L(z)}\frac{{\bf n}\cdot{\bf e}_z}{\cos\theta}\,D_pp\,dl =
- D_{ps}p_s {dS\over dz}\,.  \qquad\qquad\qquad\,
\label{eq:4.8}
\end{equation}

By substituting expressions (\ref{eq:4.6})--(\ref{eq:4.8}) into
Eq.~(\ref{eq:4.5}) and introducing the effective lifetime $\bar{\tau}_p$
to allow for both the volume and surface recombination
\[
\frac{1}{\bar{\tau}_p}=\frac{1}{\tau_p}+
2s_p\frac{\sqrt{1+ (da/dz)^2}}{a(z)}\,
\frac{p_s-p_n}{\bar{p}-p_n}\,,
\]
we arrive at the quasi-1D drift-diffusion equation for holes
injected into the $n$-region:
\[
{\partial\over\partial t}\,(\bar{p}S) +
{\partial\over\partial z}\,(\bar{p}\bar{\mu}_p \bar{E}_z S) -
{\partial^2\over\partial z^2}\,(\bar{D}_p\bar{p}S) +
{\partial\over\partial z} \left(D_{ps}p_s{dS\over dz} \right)
\]
\begin{equation}  \qquad\qquad\qquad\quad
= - \frac{\,\bar{p}-p_n}{\bar{\tau}_p}\,S.
\label{eq:4.9}
\end{equation}

Similarly, from the initial equations (\ref{eq:4.2}) and (\ref{eq:4.4})
we obtain the quasi-1D drift-diffusion equation for electrons injected
into the $p$-region:
\[
{\partial\over\partial t}\,(\bar{n}S) +
{\partial\over\partial z}\,(\bar{n}\bar{\mu}_n\bar{E}_zS) -
{\partial^2\over\partial z^2}\,(\bar{D}_n\bar{n}S) +
{\partial\over\partial z} \left(D_{ns}n_s{dS\over dz} \right)
\]
\begin{equation}  \qquad\qquad\qquad\quad
= - \frac{\bar{n}-n_p}{\bar{\tau}_n}\,S.
\label{eq:4.10}
\end{equation}

All the average quantities $\bar{\mu}_{p(n)}, \bar D_{p(n)}$, and
$\bar{\tau}_{p(n)}$ will be considered as phenomenologically given
parameters with dropping the bar sign over them~for~simplicity.

Quasi-one-dimensional equations (\ref{eq:4.9}) and (\ref{eq:4.10})
are of the general form applicable for both the $PiN$- and
PN-diodes. For low level of injection in the PN-diodes
we can assume $\bar{E}_z\!=0$ for diode base so that
Eqs.~(\ref{eq:4.9}) and (\ref{eq:4.10}) take the simplified form:
\[
{\partial\over\partial t}\,(\bar{p}S) -
{\partial\over\partial z}\, \Bigl( S{\partial\over\partial z}\,
(D_p\bar{p}) \Bigr) -
{\partial\over\partial z} \left[ (D_p\bar{p} - D_{ps}p_s)
{dS\over dz} \right]
\]
\begin{equation}  \qquad\qquad\qquad\quad
= - \frac{\,\bar{p}-p_n}{{\tau}_p}\,,
\label{eq:4.11}
\end{equation}
\[
{\partial\over\partial t}\,(\bar{n}S) -
{\partial\over\partial z}\, \Bigl( S{\partial\over\partial z}\,
(D_n\bar{n}) \Bigr) -
{\partial\over\partial z} \left[ (D_n\bar{n} - D_{ns}n_s)
{dS\over dz} \right]
\]
\begin{equation}  \qquad\qquad\qquad\quad
= - \frac{\,\bar{n}- n_p}{{\tau}_n}\,.
\label{eq:4.12}
\end{equation}

Not having a specific knowledge of surface properties, we shall assume
that ${D_{ps}p_s=D_p\bar{p}}$\, and ${D_{ns}n_s=D_n\bar{n}}$.
Then, from equations (\ref{eq:4.11}) and (\ref{eq:4.12}) follows the
{\em quasi-one-dimensional diffusion equations\/} for the excess
concentrations of holes, ${\Delta\bar{p}(z,t)=\bar{p}(z,t)-p_n}$, and
electrons, ${\Delta\bar{n}(z,t)=\bar{n}(z,t)-n_p}$, injected into
the appropriate neutral parts of the PN-diode~\cite{BS2007a}:
\begin{equation}
\biggl( 1+ {\tau}_p\,
{\partial\over\partial t} \biggr) \Delta\bar{p} -
L_p^2\, {\partial^2\Delta\bar{p}\over\partial z^2} -
L_p^2\, {\partial \ln S\over\partial z}\,
{\partial\Delta\bar{p}\over\partial z} = 0,
\label{eq:4.13}
\end{equation}
\begin{equation}
\biggl( 1+ {\tau}_n\,
{\partial\over\partial t} \biggr) \Delta\bar{n} -
L_n^2\, {\partial^2\Delta\bar{n}\over\partial z^2} -
L_n^2\, {\partial \ln S\over\partial z}\,
{\partial\Delta\bar{n}\over\partial z} = 0,
\label{eq:4.14}
\end{equation}
where $L_{p(n)} =\sqrt{D_{p(n)}{\tau}_{p(n)}}$ is the diffusion
length for holes (electrons). The additional last term involving 
$\partial \ln S/\partial z\equiv S'(z)/S(z)$ 
takes into account the cross-sectional nonuniformity.

\section{Spectral Solution of Quasi-One-Dimensional Diffusion Equations}
\label{sec:3}

In general case, the voltage applied to the PN-junction consists of
the DC bias voltage $V_0$ and the AC harmonic signal $V_\sim\cos\omega t$:
\begin{equation}
v(t)= V_0 + V_\sim \cos\omega t \,\equiv
 V_0 + \bigl[ V_1(\omega) e^{i\omega t}+ c.c. \bigr].
\label{eq:4.15}
\end{equation}

Nonlinearity of electronic processes in the PN-junction produces the
frequency harmonics $k\omega$ so that solutions of Eqs.~(\ref{eq:4.13})
and (\ref{eq:4.14}) have the form of Fourier series:
\begin{equation}
\Delta \bar{p}(z,t)=\!\!
\sum_{k\,=-\infty}^{\infty} \!\Delta\bar{p}_k(z) \,e^{ik\omega t},
\label{eq:4.16}
\end{equation}
\begin{equation}
\Delta \bar{n}(z,t)=\!\!
\sum_{k\,=-\infty}^{\infty} \!\Delta \bar{n}_k(z) \,e^{ik\omega t}.
\label{eq:4.17}
\end{equation}

Real values of $\Delta \bar{p}(z,t)$ and $\Delta \bar{n}(z,t)$ are
provided with the following relations for the complex amplitudes:\,
$\Delta\bar{p}_k=\Delta\bar{p}_{-k}^*$ \,and\,
$\Delta\bar{n}_k= \Delta\bar{n}_{-k}^*$.

Substitution of the required solutions (\ref{eq:4.16}) and
(\ref{eq:4.17}) into Eqs.~(\ref{eq:4.13}) and (\ref{eq:4.14}) with
regard for the orthogonality of harmonics reduces to the following
equations for the harmonic amplitudes~\cite{BS2007a}:
\begin{equation}
{d^2{\Delta \bar{p}_k}\over dz^2} +
{d\ln S\over dz}\,{d{\Delta \bar{p}_k}\over dz} -
\frac{\Delta \bar{p}_k}{L_{pk}^2}=0,  \,
\label{eq:4.18}
\end{equation}
\begin{equation}
{d^2{\Delta \bar{n}_k}\over dz^2} +
{d\ln S\over dz}\,{d{\Delta \bar{n}_k}\over dz} -
\frac{\Delta \bar{n}_k}{L_{nk}^2}=0,
\label{eq:4.19}
\end{equation}
where
\[
L_{pk}=\frac{L_p}{\sqrt{1+ik\omega\tau_p}}
\,\,\quad\mbox{and}\quad\,\,
L_{nk}=\frac{L_n}{\sqrt{1+ik\omega\tau_n}}\,.
\]

As an analytical approximation to a mesa-like structure, the special 
case of exponential change of the cross section
$S(z)=S_0\exp(2\alpha z)$ is considered, Eqs.~(\ref{eq:4.18}) and 
(\ref{eq:4.19}) turn into the linear equations~\cite{BS2007a}
\begin{equation}
{d^2{\Delta \bar{p}_k}\over dz^2} +
2\alpha\,{d{\Delta \bar{p}_k}\over dz} -
\frac{\Delta \bar{p}_k}{L_{pk}^2}=0,
\label{eq:4.20}
\end{equation}
\begin{equation}
{d^2{\Delta \bar{n}_k}\over dz^2} +
2\alpha\,{d{\Delta \bar{n}_k}\over dz} -
\frac{\Delta \bar{n}_k}{L_{nk}^2}=0.
\label{eq:4.21}
\end{equation}


Representing the desired solution in the form of $e^{-\lambda z}\!$,
we obtain the following characteristic equation for Eqs.~(\ref{eq:4.20})
and (\ref{eq:4.21}):
\begin{equation}
\lambda^2 - 2\alpha\lambda-
{L_{k}^{-2}}= 0\,, \,\,\quad \mbox{where} \,\,\,
L_k=\left\{\!
\begin{array}{ll}
L_{pk} \,\quad \mbox{for\,\,holes}, \\[2mm]
L_{nk} \,\quad \mbox{for electrons}.
\end{array} \right.
\label{eq:4.22}
\end{equation}
The complex roots of Eq.~(\ref{eq:4.22}) can be written as follows
\begin{equation}
\lambda_{1,2}=\left\{\!
\begin{array}{ll}
\pm\,\Lambda_{pk}^\pm\,/L_p
\,\,\,\quad\mbox{for\,\,the}\,n\mbox{-region }  (z\geq W_n), \\[2mm]
\pm\,\Lambda_{nk}^\pm\,/L_n
\,\:\quad\mbox{for\,\,the}\,p\mbox{-region }  (z\leq -W_p),
\end{array} \right.
\label{eq:4.23}
\end{equation}
where we have introduced the following notations:\\[1mm]
$\bullet$\,\,for holes injected into the $n$-region ($z\geq W_n$)
\[
\Lambda_{pk}^{\pm}= (a_{pk}^\alpha A_p^\alpha \pm\alpha L_p)+
ib_{pk}^\alpha A_p^\alpha , \qquad\,\,
A_p^\alpha = \sqrt{1+(\alpha L_p)^2}\,,
\]
\begin{equation} \qquad\qquad
a_{pk}^\alpha= \frac{1}{\sqrt{2}}\,\sqrt{1+\Theta_{pk}^{\alpha}}\,,\qquad\,\,\,
b_{pk}^\alpha= \frac{k\omega\tau_p^{\alpha}}{2a_{pk}^\alpha}\,,
\label{eq:4.24}
\end{equation}
\[ \qquad\qquad
\Theta_{pk}^{\alpha}=\sqrt{1+(k\omega\tau_p^{\alpha})^2}\,, \qquad\,\,
\tau_p^{\alpha}=\frac{\tau_p}{1+(\alpha L_p)^2}\,;
\]
$\bullet$\,\,for electrons injected into the $p$-region ($z\leq -W_p$)
\[
\Lambda_{nk}^{\pm}= (a_{nk}^\alpha A_n^\alpha \pm\alpha L_n)+
ib_{nk}^\alpha A_n^\alpha , \qquad\,\,
A_n^\alpha = \sqrt{1+(\alpha L_n)^2
   \vphantom{(\alpha L_p)^2}}\,,
\]
\begin{equation} \qquad\qquad\;\;
a_{nk}^\alpha= \frac{1}{\sqrt{2}}\,\sqrt{1+ \Theta_{nk}^\alpha
                         \vphantom{\Theta_{pk}^\alpha}}\,,  \qquad\;\;
b_{nk}^\alpha= \frac{k\omega\tau_n^{\alpha}}{2a_{nk}^\alpha}\,,
\label{eq:4.25}
\end{equation}
\[ \qquad\qquad\,
\Theta_{nk}^{\alpha}=\sqrt{1+(k\omega\tau_n^\alpha)^2
                           \vphantom{\tau_p}}\,, \qquad\,\,
\tau_n^{\alpha}=\frac{\tau_n}{1+(\alpha L_n)^2}\,.
\]

The general solutions of equations (\ref{eq:4.20}) and (\ref{eq:4.21})
have the following form: \\[2mm]
$\bullet$\,\,for holes injected into the $n$-region ($z\geq W_n$)
\begin{equation}
\Delta \bar{p}_k(z)=
C_{pk}^+ \exp \biggl(\! -\Lambda_{pk}^+\frac{z-W_n}{L_p} \biggr) +\,
C_{pk}^- \exp \biggl( \Lambda_{pk}^-\frac{z-W_n}{L_p} \biggr) ;
\label{eq:4.26}
\end{equation}
$\bullet$\,\,for electrons injected into the $p$-region ($z\leq -W_p$)
\begin{equation}
\Delta \bar{n}_k(z)=
C_{nk}^+ \exp \biggl(\! -\Lambda_{nk}^+\frac{z+W_p}{L_n} \biggr) +\,
C_{nk}^- \exp \biggl( \Lambda_{nk}^-\frac{z+W_p}{L_n} \biggr) .
\label{eq:4.27}
\end{equation}

In accordance with notations (\ref{eq:4.24}) and (\ref{eq:4.25}), always
Re\,$\Lambda_{pk}^{\pm}=a_{pk}^\alpha A_p^\alpha \pm\alpha L_p>0$ and
Re\,${\Lambda_{nk}^{\pm}=a_{nk}^\alpha A_n^\alpha \pm\alpha L_n>0}$
so that for the PN-diodes with thick bases we can
assume $C_{pk}^- =0$ and $C_{nk}^+ =0$, which eliminates the necessity
for boundary conditions on ohmic contacts. Taking into account formulas
(\ref{eq:4.26}) and (\ref{eq:4.27}) with $C_{pk}^-= C_{nk}^+=0$,
the general solutions (\ref{eq:4.16}) and (\ref{eq:4.17}) of
Eqs.~(\ref{eq:4.13}) and (\ref{eq:4.14}) are written in the form
\begin{equation}
\Delta \bar{p}(z,t)=\!\! \sum_{k\,=-\infty}^{\infty}\!
C_{pk}^+ \exp \biggl(\! -\Lambda_{pk}^+\frac{z-W_n}{L_p} \biggr)
e^{ik\omega t} \,,
\label{eq:4.28}
\end{equation}
\begin{equation}
\Delta \bar{n}(z,t)=\!\! \sum _{k\,=-\infty}^{\infty}\!
C_{nk}^-\exp \biggl( \Lambda_{nk}^-\frac{z+W_p}{L_n} \biggr)
e^{ik\omega t} \,.
\label{eq:4.29}
\end{equation}

These expressions provide $\Delta\bar{p}(z,t)\to 0$ as $z\to\!\infty$
and $\Delta\bar{n}(z,t)\to 0$ as $z\to\!-\infty$ because of
Re\,$\Lambda_{pk}^+>0$ and Re\,$\Lambda_{nk}^->0$.

The constants $C_{pk}^+$ and $C_{nk}^-$ appearing in Eqs.~(\ref{eq:4.28})
and (\ref{eq:4.29}) can be found from the conventional injection boundary
conditions~\cite{SMS1981}:
\begin{equation}
\Delta \bar{p}(W_n,t)= p_n f(t) \quad\quad \mbox{for}\,\, z=W_n, \,\,\,
\label{eq:4.30}   \\[0.5mm]
\end{equation}
\begin{equation}
\Delta \bar{n}(-W_p,t)= n_p f(t) \quad\,\mbox{for}\,\, z=-W_p,
\label{eq:4.31}
\end{equation}
where for the applied voltage $v(t)$ of the form (\ref{eq:4.15}) we have
introduced the function
\begin{equation}
f(t)=\exp \biggl( \frac{qv(t)}{\kappa  T} \biggr) - 1.
\label{eq:4.32}
\end{equation}

Substitution of expressions (\ref{eq:4.28}) and (\ref{eq:4.29})
into the boundary conditions (\ref{eq:4.30}) and (\ref{eq:4.31}) yields
\begin{eqnarray}
&& \sum_{k\,=-\infty}^{\infty} \!C_{pk}^+
\,e^{ik\omega t} = p_n f(t) , \nonumber
\\[-1.5mm] &\label{eq:4.33}& \\[-1.5mm]
&& \sum_{k\,=-\infty}^{\infty} \!C_{nk}^-
\,e^{ik\omega t} = n_p f(t).  \nonumber  \qquad\quad
\end{eqnarray}

By using the orthogonality property of harmonics in expansions
(\ref{eq:4.33}) it is easy to get the desired constants
\begin{equation}
C_{pk}^+ = p_n F_k \qquad\mbox{and}\qquad C_{nk}^-= n_p F_k,
\label{eq:4.34}
\end{equation}
where $F_k$ is the Fourier amplitude of the $k$th harmonic for the
function $f(t)$ given by formula (\ref{eq:4.32}), that is
\begin{equation}
F_k = \frac{1}{2\pi} \int\limits_{-\pi}^{\,\pi}
f(t) e^{-ik\omega t} \,d\omega t .
\label{eq:4.35}
\end{equation}
Substitution of the function (\ref{eq:4.32}) into formula (\ref{eq:4.35}) gives
\begin{equation}
F_0 = I_0(\beta V_{\sim})\exp (\beta V_0) - 1
\qquad\quad \,\mbox{for} \quad k=0,
\label{eq:4.36}
\end{equation}
\begin{equation}
F_k= F_{-k} = I_k(\beta V_{\sim})\exp (\beta V_0)
\qquad \mbox{for} \quad k\neq 0,
\label{eq:4.37}
\end{equation}
where we have used the modified Bessel functions of the first kind
of order $k$ ($k=0,\,\pm 1,\,\pm 2, \ldots$) having the following integral
representation (see formula 8.431.5 in Ref.~\cite{GR1980}):
\begin{equation}
I_k(\beta V_{\sim})=\frac{1}{\pi}
\int\limits_0^{\,\pi} e^{ \beta V_{\sim}\cos\omega t}
\cos k\omega t \:d\omega t.
\label{eq:4.38}
\end{equation}
This function depends on $\beta V_{\sim}$, where $V_{\sim}$ is an
amplitude of the signal applied to the PN-junction \,and\,
$\beta = q/\kappa T$.

Thus, with allowing for Eq.~(\ref{eq:4.34}) the general solutions
(\ref{eq:4.28}) and (\ref{eq:4.29}) of the diffusion equations
(\ref{eq:4.13}) and (\ref{eq:4.14}) take the final form of spectral
expansions:
\begin{equation}
\Delta \bar{p}(z,t)=\, p_n\! \sum_{k\,=-\infty}^{\infty}\!
F_k \exp \biggl(\! -\Lambda_{pk}^+\frac{z-W_n}{L_p} \biggr)
e^{ik\omega t} ,
\label{eq:4.39}
\end{equation}
\begin{equation}
\Delta \bar{n}(z,t)=\, n_p\! \sum _{k\,=-\infty}^{\infty}\!
F_k \exp \biggl( \Lambda_{nk}^-\frac{z+W_p}{L_n} \biggr)
e^{ik\omega t} .
\label{eq:4.40}
\end{equation}

These expressions allow us to obtain the spectral composition of
the current flowing through the external circuit connected to the
PN-diode.

\section{External Circuit Current for Semiconductor Diode with
          Nonuniform Cross Section}
\label{sec:4}

The initial equation to derive an expression for the diode current is the
law of total current conservation:
\begin{equation}
\nabla\cdot \biggl( {\bf j}_p +{\bf j}_n +
\epsilon\,{\partial{\bf E}\over\partial t} \biggr) = 0
\label{eq:4.41}
\end{equation}
following from Maxwell's equation $\nabla\!\times{\bf H}=
{\bf j}_p+{\bf j}_n+\epsilon\,\partial{\bf E}/\partial t$.

The transverse averaging technique applied to Eq.~(\ref{eq:4.41}) gives
\[
{\partial\over\partial z}
\biggl[ \biggl( \bar{\!j}_{pz}+ \,\bar{\!j}_{nz} +
\epsilon\, {\partial\bar{E}_z\over\partial t} \biggr)S \biggr]
\qquad\qquad\qquad\qquad\qquad\quad
\]
\begin{equation}  \qquad\qquad\quad
+\! \oint\limits_{L(z)}\!\frac{{\bf n}\cdot{\bf j}}{\cos\theta}\,dl\,+\,
{\partial\over\partial t}
\oint\limits_{L(z)}\!\frac{{\bf n}\cdot\epsilon {\bf E}}
{\cos\theta}\,dl=0 .
\label{eq:4.42}
\end{equation}

The similar equation outside the semiconductor, where currents are absent and
${\nabla\cdot(\epsilon^o{\bf E}^o)=0}$, has the following form:
\begin{equation}
{\partial\over\partial z}
\biggl( \epsilon^o {\partial\bar{E}_z^o\over\partial t}\,S^o \biggr) -
{\partial\over\partial t} \oint\limits_{L(z)}\!
\frac{{\bf n}\cdot\epsilon^o{\bf E}^o}{\cos\theta} \,dl=0,
\label{eq:4.43}
\end{equation}
where $S^o$ is an effective localization area of the fringe outside
field ${\bf E}^o$ such that usually $S^o\!\ll S$ and
$\epsilon^o\!<\epsilon$. The addition of Eqs.~(\ref{eq:4.42}) and
(\ref{eq:4.43}) gives
\[
{\partial\over\partial z}
\biggl[ \Bigl(\, \bar{\!j}_{pz} +\, \bar{\!j}_{nz} \Bigr)S +
{\partial\over\partial t}
\left( \epsilon \bar{E}_zS +
\epsilon^o \bar{E}_z^o S^o\right) \biggr] \qquad\qquad\quad\,
\]
\begin{equation}  \qquad\,
+\! \oint\limits_{L(z)}\!\frac{{\bf n}\cdot
{\bf j}}{\cos\theta}\,dl \,+\,
{\partial\over\partial t} \oint\limits_{L(z)}\!\frac{{\bf n}\cdot
(\epsilon{\bf E} - \epsilon^o{\bf E}^o) }{\cos\theta}\,dl=0 .
\label{eq:4.44}
\end{equation}

If a semiconductor surface contains traps with the charge density
$\rho_s$, there are the following boundary conditions on the contour
$L(z)$~\cite{Barybin1986}:
\begin{equation}
{\bf n}\cdot\epsilon^o{\bf E}^o =
{\bf n}\cdot\epsilon\,{\bf E} + \rho_s
\,\quad\mbox{and}\quad\;
{\partial\rho_s\over\partial t}= {\bf n}\cdot{\bf j}\,.
\label{eq:4.45}
\end{equation}

In this case, two contour integrals in Eq.~(\ref{eq:4.44}) cancel each
other and with allowing for inequality $|\epsilon^o\bar{E}^o_z S^o|\ll
|\epsilon\bar{E}_z S|$ formula
(\ref{eq:4.44}) takes the form
\begin{equation}
{\partial\over\partial z} \biggl[ \biggl(
\bar{\!j}_{pz}(z,t) +\,\bar{\!j}_{nz}(z,t) +
\epsilon\,{\partial{\bar E_z(z,t)}\over\partial t}\biggr)S(z) \biggr]=0.
\label{eq:4.46}
\end{equation}

The quantity in brackets of Eq.~(\ref{eq:4.46}), being independent of $z$,
defines the {\em external circuit current\/} equal to
\begin{equation}
J(t)= q(\mu_p\bar p+ \mu_n\bar n)\bar E_zS
- q\biggl(\! D_p{\partial{\bar p}\over\partial z} -
D_n {\partial{\bar n}\over\partial z} \biggr) S +
\epsilon\, {\partial{\bar E_z}\over\partial t}\,S .
\label{eq:4.47}
\end{equation}
Here we have used expressions (\ref{eq:4.6}) and (\ref{eq:4.8})
for the average hole current\, $\bar{\!j}_{pz}S$ and the similar
expressions for the average electron current\, $\bar{\!j}_{nz}S$
(with dropping the bar sign over $\bar\mu_{p(n)}$ and $\bar D_{p(n)}$).
All the terms on the right of Eq.~(\ref{eq:4.47}) depend on both
$z$ and $t$ but taken together at any cross section $S(z)$ they yield
the external circuit current $J(t)$ as a function of only time.

Restricting our consideration to the
PN-diodes with low injection, we can assume $\bar{E}_z=0$ in
neutral parts of the $p$- and $n$-regions~\cite{SMS1981}. Then the external
circuit current (\ref{eq:4.47}) is determined only by the averaged
diffusion currents taken at any cross section $S(z)$, for example,
at $z=W_n$:
$$ J(t)=
-qD_p {\partial\bar p(z,t)\over\partial z} \bigg|_{z=W_n}\!\!S(W_n)\,+\,
qD_n {\partial\bar n(z,t)\over\partial z} \bigg|_{z=W_n}\!\!S(W_n)
$$
\begin{equation} \qquad
\equiv\,\bar{\!j}_{pz}(W_n,t)S(W_n) +\,\bar{\!j}_{nz}(W_n,t)S(W_n). \quad\,
\label{eq:4.48}
\end{equation}

Neglecting recombination processes inside the PN-junction, which is
true if $W=W_n+W_p\ll L_p$ and $L_n$~\cite{SMS1981}, we can write
\begin{equation}
\bar{\!j}_{nz}(W_n,t)S(W_n)=\, \bar{\!j}_{nz}(-W_p,t)S(-W_p).
\label{eq:4.49}
\end{equation}

Substitution of relation (\ref{eq:4.49}) into Eq.~(\ref{eq:4.48}) gives
the external circuit current (cf.~Eq.~(31) in Ref.~\cite{BS2007a})
\[
J(t)=\,\bar{\!j}_{pz}(W_n,t)S(W_n)+\,\bar{\!j}_{nz}(-W_p,t)S(-W_p)  \,
\]
\begin{equation}
= -\,qD_p {\partial\Delta\bar p(z,t)\over\partial z}
\bigg|_{\,z\,=\,W_n}\!\!S(W_n) +\,
qD_n {\partial\Delta\bar n(z,t)\over\partial z}
\bigg|_{\!\,z\,=\,-W_p}\!S(-W_p),
\label{eq:4.50}
\end{equation}
where $\Delta\bar p =\bar p -p_n$ and $\Delta\bar n =\bar n -n_p$
are the excess concentrations of injected carriers determined by
formulas (\ref{eq:4.39}) and (\ref{eq:4.40}). Inserting these
formulas into expression (\ref{eq:4.50}), we finally obtain the
spectral representation for the external circuit current:
\begin{equation}
J(t)=J_{s,p}\! \sum_{k\,=-\infty}^{\infty}\!
F_k\Lambda^+_{pk}\,{\rm e}^{ik\omega t} \,+\,
J_{s,n}\! \sum_{k\,=-\infty}^{\infty}\!
F_k\Lambda^-_{nk}\,{\rm e}^{ik\omega t}.
\label{eq:4.51}
\end{equation}
Here, the hole and electron contributions into the saturation current
of a thick PN-junction are defined, as it is generally
accepted~\cite{SMS1981}, in the form
\begin{equation}
J_{s,p} =\frac{qD_p p_n}{L_p}\,S_n  \,\quad\mbox{and}\quad
J_{s,n} =\frac{qD_n n_p}{L_n}\,S_p,
\label{eq:4.52}
\end{equation}
where
\[S_n\equiv S(W_n)=S_0\exp(2\alpha W_n), \quad\, \]
\[S_p\equiv S(-W_p)=S_0\exp(-2\alpha W_p).\]

Expression (\ref{eq:4.51}) contains all the spectral components
of the external circuit current including the DC current for $k=0$
and the AC current for $k=\pm\,1$ which are of most interest for
our further consideration.

\section{Static Current--Voltage Characteristic of PN-Diode
            with Nonuniform Cross Section}
\label{sec:5}

The term in series (\ref{eq:4.51}) numbered by $k=0$ corresponds to the DC
current $J_0$, which by using (\ref{eq:4.36}) can be written in the form
of the {\it static current--voltage characteristic}:
\begin{equation}
J_0(V_0,V_{\sim})=
J_s \Bigl[\!\: I_0(\beta V_{\sim})\,e^{\beta V_0} -\!1 \!\:\Bigr].
\label{eq:4.53}
\end{equation}
Here we have introduced the saturation current for a nonuniform diode
\begin{equation}
J_s = J_{s,p}\!\:r_{p0}^{\alpha} + J_{s,n}\!\:r_{n0}^{\alpha}
\label{eq:4.54}
\end{equation}
and, in accordance with expressions (\ref{eq:4.24}) and (\ref{eq:4.25})
for $k=0$, have used the following notation:
\begin{eqnarray}
&& r_{p0}^{\alpha}\equiv\Lambda^+_{p0}= A_p^\alpha + \alpha L_p =
\sqrt{1+{(\alpha L_p)}^2} +\alpha L_p,\,
\nonumber \\[-1.5mm]&\label{eq:4.55}& \\[-1.5mm]
&& r_{n0}^{\alpha}\equiv\Lambda^-_{n0}= A_n^\alpha - \alpha L_n =
\sqrt{1+{(\alpha L_n)}^2}- \alpha L_n.
\nonumber  \qquad\quad
\end{eqnarray}

For the uniform $p-n$-junction (with $\alpha=0$) we have
$r_{p0}^{\alpha}=r_{n0}^{\alpha}=1$ and $S_n=S_p=S_0$ so that
the static current-voltage characteristic
retains the form (\ref{eq:4.53}) with the usual saturation current
\begin{equation}
J_s = J_{s,p}^0 + J_{s,n}^0 \equiv
\frac{qD_p p_n}{L_p}\,S_0 + \frac{qD_n n_p}{L_n}\,S_0,
\label{eq:4.56}
\end{equation}
where as before superscript 0 marks the cross-sectional uniformity
(when $\alpha =0$ and $S_0=$ constant).

The modified Bessel function $I_0(\beta V_{\sim})$ appearing in
Eq.~(\ref{eq:4.53}) is completely the same as that obtained in our
paper~\cite{BS2007a} and it distinguishes our expression (\ref{eq:4.53})
from the similar formula given in the known literature on semiconductor
electronics~\cite{SMS1981}. Coincidence between them occurs only for
such small signals that $V_{\sim}\!\ll\!\kappa T/q$ and
$I_0(\beta V_{\sim})\simeq 1$. The function ${I_0(\beta V_{\sim})\geq 1}$
reflects the effect of signal rectification, which provides the contribution
into the DC current from a signal and results in upward shifts of curves
$J_0(V_0)$ with increasing the signal amplitude $V_{\sim}$.

\section{Dynamic Impedance of PN-Diode with
                        Nonuniform Cross Section}
\label{sec:6}

The first harmonic of the external current in the general expression
(\ref{eq:4.51}) corresponds to terms numbered by $k=\pm\,1$ and equals
\begin{equation}
J_1(t)=J_1(\omega) e^{i\omega t} +c.c.,  \quad\,\,
\label{eq:4.57}
\end{equation}
where $J_1(\omega) = F_1\, \bigl[ J_{s,p}\,\Lambda^+_{p1}(\omega) +
J_{s,n}\,\Lambda^-_{n1}(\omega) \bigr]$. The quantities $\Lambda^+_{p1}$,\, $ \Lambda^-_{n1}$, and $F_1$ are
respectively defined by formulas (\ref{eq:4.24}), (\ref{eq:4.25}), and
(\ref{eq:4.37}) for $k=1$.

Expressions (\ref{eq:4.15}) with $V_\sim\!=2V_1(\omega)$ and (\ref{eq:4.57})
allow one to introduce the {\em dynamic admittance\/} of the PN-diode
as a function of frequency:
\begin{equation}
Y(\omega)=\frac{J_1(\omega)}{V_1(\omega)}\equiv G_d(\omega)+
i\omega C_d(\omega).
\label{eq:4.58}
\end{equation}

The {\it dynamic\/} ({\it diffusion\/}) {\it conductance\/} $G_d$ and the
{\it dynamic\/} ({\it diffusion\/}) {\it capacitance\/} $C_d$ are defined
in a customary way~\cite{SMS1981}. After substituting (\ref{eq:4.57}) into
Eq.~(\ref{eq:4.58}) and some transformations with regard for (\ref{eq:4.54}),
we obtain
\begin{eqnarray}
G_d =
gG_0\,\frac{J_{s,p}\!\:r_{p1}^{\alpha} + J_{s,n}\!\:r_{n1}^{\alpha}}
{J_{s,p}\!\:r_{p0}^{\alpha} + J_{s,n}\!\:r_{n0}^{\alpha}} \equiv
gG_0 \left( r_{p1}^{\alpha}\frac{J_{s,p}}{J_s} +
r_{n1}^{\alpha}\frac{J_{s,n}}{J_s} \right),  \qquad
\label{eq:4.59}
\end{eqnarray}
\begin{eqnarray}
C_d =
\frac{gG_0}{\omega}\, \frac{J_{s,p}\!\:q_{p1}^{\alpha} +
J_{s,n}\!\:q_{n1}^{\alpha}}
{J_{s,p}\!\:r_{p0}^{\alpha} + J_{s,n}\!\:r_{n0}^{\alpha}} \equiv
\frac{gG_0}{2} \left( q_{p1}^{\alpha}\frac{J_{s,p}\tau_p}{J_s} +
q_{n1}^{\alpha}\frac{J_{s,n}\tau_n}{J_s} \right).
\label{eq:4.60}
\end{eqnarray}
Here we have introduced the new quantities:
\begin{equation}
r_{p1}^{\alpha}=\,a_{p1}^\alpha A_p^\alpha + \alpha L_p\,, \qquad
r_{n1}^{\alpha}=\,a_{n1}^\alpha A_n^\alpha - \alpha L_n\,,
\label{eq:4.61}
\end{equation}
\begin{equation}
q_{p1}^{\alpha}= \frac{1}{a_{p1}^\alpha A_p^\alpha} \,,
\qquad\qquad\quad
q_{n1}^{\alpha}= \frac{1}{a_{p1}^\alpha A_p^\alpha} \,,  \qquad\quad
\label{eq:4.62}
\end{equation}
and, in accordance with Eqs.~(\ref{eq:4.24}) and (\ref{eq:4.25})
for $k=1$, used the following notation:
\begin{eqnarray}
a_{p1}^\alpha= \frac{1}{\sqrt{2}}\,\sqrt{1+\Theta^{\alpha}_{p1}}\:,\qquad\,\,
a_{n1}^\alpha= \frac{1}{\sqrt{2}}\,\sqrt{1+\Theta^{\alpha}_{n1}
                            \vphantom{\Theta^{\alpha}_{p1}}}\:, \quad &&
\nonumber \\[-2mm]&\label{eq:4.63}& \\[-2mm]
\Theta^{\alpha}_{p1}=\sqrt{1+{(\omega\tau^{\alpha}_p)}^2}, \qquad\,\,\,
\Theta^{\alpha}_{n1}=\sqrt{1+{(\omega\tau^{\alpha}_n)}^2}.      \:\quad &&
\nonumber
\end{eqnarray}

Formulas (\ref{eq:4.59}) and (\ref{eq:4.60}) contain the differential
conductance $G_0$ of the static current--voltage characteristic
$J_0(V_0, V_\sim)$ defined as
\begin{equation}
G_0= {\partial J_0(V_0, V_\sim)\over\partial V_0}\,.
\label{eq:4.64}
\end{equation}

For the static characteristic of form (\ref{eq:4.53}), the differential
conductance (\ref{eq:4.64}) depends on both the bias voltage $V_0$ and
the signal amplitude $V_\sim$:
\begin{equation}
G_0(V_0,V_{\sim}) = \frac{J_0 + J_s}{\kappa T/q} \,\equiv\,
{J_s\exp(\beta V_0)\over \kappa T/q} \,I_0(\beta V_\sim).
\label{eq:4.65}
\end{equation}

Formulas (\ref{eq:4.59}) and (\ref{eq:4.60}) involve a quantity
\begin{equation}
g(V_\sim) =
{g_1(V_\sim)\over I_0(\beta V_\sim)} \leq 1
\,\quad\,\mbox{with}\,\,\quad
g_1(V_\sim) = \frac{I_1(\beta V_{\sim})}{\beta V_{\sim}/2}
\label{eq:4.66}
\end{equation}
composed of the modified Bessel functions $I_0$ and $I_1$ and introduced
before in paper~\cite{BS2007a}. It depends on the signal amplitude $V_{\sim}$, so
that ${g(V_\sim)\simeq 1}$ for small signals when $V_{\sim}\ll\kappa T/q$\,
and\, $g(V_\sim)\to 0$ as a function $2/(\beta V_\sim$) when $V_\sim\!\to\infty$.

From Eqs.~(\ref{eq:4.65}) and (\ref{eq:4.66}) it follows that
\begin{equation}
gG_0 = g_1\frac{J_s\exp(\beta V_0)}{\kappa T/q} \equiv
\frac{I_1(\beta V_\sim)}{\beta V_\sim/2}\,
\frac{J_s\exp(\beta V_0)}{\kappa T/q}\,.
\label{eq:4.67}
\end{equation}

After substituting (\ref{eq:4.67}) into Eqs.~(\ref{eq:4.59})
and (\ref{eq:4.60}), they assume the following form:
\begin{equation}
G_d(\omega) = g_1 G_{d0} \,\biggl(
r_{p1}^{\alpha}(\omega) {J_{s,p}\over J_s} \!\:+\!\:
r_{n1}^{\alpha}(\omega) {J_{s,n}\over J_s} \biggr),
\label{eq:4.68}
\end{equation}
\begin{equation}
C_d(\omega) = g_1 C_{d0} \,\biggl(
q_{p1}^{\alpha}(\omega) {Q_p\over Q} \!\:+\!\:
q_{n1}^{\alpha}(\omega) {Q_n\over Q} \biggr), \;\,
\label{eq:4.69}
\end{equation}
where, following to paper~\cite{BS2007a}, we have introduced the charges
$Q_p\equiv J_{s,p}\tau_p = qL_p p_n S_n$,
$Q_n\equiv J_{s,n}\tau_n = qL_n n_p S_p$, and similar to
Eq.~(\ref{eq:4.54})
\begin{equation}
Q = Q_p\!\:r_{p0}^{\alpha} + Q_n\!\:r_{n0}^{\alpha}.
\label{eq:4.70}
\end{equation}

The quantities $G_{d0}$ and $C_{d0}$ appearing in
Eqs.~(\ref{eq:4.68}) and (\ref{eq:4.69}) are defined, by analogy
with those in paper~\cite{BS2007a},\, as 
\begin{eqnarray}
G_{d0}(V_0)
&=&
{J_s\exp(\beta V_0)\over\kappa T/q} =
{J_{s,p}\!\:r_{p0}^{\alpha} +
J_{s,n}\!\:r_{n0}^{\alpha}\over\kappa T/q}\;e^{\beta V_0}
\nonumber\\[1.5mm]
&\equiv&
{q\over\kappa T} \biggl( \frac{qp_n D_p}{L_p}\,S_n\!\:r_{p0}^{\alpha} +
\frac{qn_p D_n}{L_n}\,S_p\!\:r_{n0}^{\alpha} \biggr) e^{qV_0/\kappa T}\!,
\qquad\qquad
\label{eq:4.71}
\end{eqnarray}
\begin{eqnarray}
C_{d0}(V_0)
&=&
{Q\exp(\beta V_0)\over 2\kappa T/q} =
{Q_p\!\:r_{p0}^{\alpha} +
Q_n\!\:r_{n0}^{\alpha}\over 2\kappa T/q}\;e^{\beta V_0}
\nonumber\\[1.5mm]
&\equiv&
{q\over\kappa T} \biggl( \frac{qp_n L_p}{2}\,S_n\!\:r_{p0}^{\alpha} +
\frac{qn_p L_n}{2}\,S_p\!\:r_{n0}^{\alpha} \biggr)e^{qV_0/\kappa T}\!.
\qquad\qquad
\label{eq:4.72}
\end{eqnarray}

An essential simplification of Eqs.~(\ref{eq:4.68})--(\ref{eq:4.69})
and (\ref{eq:4.71})--(\ref{eq:4.72}) occurs in the case of the
{\em one-sided\/} $P^+N$-junction with highly doped
emitter when $p_{n}\gg n_{p}$, $W_p\ll W_n$, $J_{s,n}\!\ll J_{s,p}$,
$Q_n\!\ll Q_p$,\, so that Eqs.~(\ref{eq:4.54}) and (\ref{eq:4.70}) yield
$J_s\simeq J_{s,p}\!\:r_{p0}^{\alpha}$ \,and\, $Q\simeq Q_p\!\:r_{p0}^{\alpha}$.
Then, the quantities (\ref{eq:4.71}) and (\ref{eq:4.72}) can be written
in the simplified form:
\begin{equation}
G_{d0}(V_0)\simeq G_{d0}^0(V_0)\!\:r_{p0}^{\alpha}\,e^{2\alpha W_n},
\label{eq:4.73}
\end{equation}
\begin{equation}
C_{d0}(V_0)\simeq C_{d0}^0(V_0)\!\:r_{p0}^{\alpha}\,e^{2\alpha W_n},
\label{eq:4.74}
\end{equation}
where the newly introduced quantities (marked with superscript 0)
\begin{equation}
G_{d0}^0(V_0) = \frac{J_{s,p}^0\exp(\beta V_0)}{\kappa T/q} \equiv
\frac{qS_0}{\kappa T}\,\frac{qp_n D_p}{L_p}\,e^{qV_0/\kappa T}\!, \;\;
\label{eq:4.75}
\end{equation}
\begin{equation}
C_{d0}^0(V_0) = \frac{Q_p^0\exp(\beta V_0)}{2\kappa T/q} \equiv
\frac{qS_0}{\kappa T}\,\frac{qp_n L_p}{2}\,e^{qV_0/\kappa T}\!,
\label{eq:4.76}
\end{equation}
correspond, as before, to the cross-sectionally uniform structures
(with $\alpha =0$ and $S_0=$ constant).

Substitution of Eqs.~(\ref{eq:4.73}) and (\ref{eq:4.74}) into
expressions (\ref{eq:4.68}) and (\ref{eq:4.69}) with $J_{s,n}\simeq 0$
and $Q_n\simeq 0$ converts them into the form appropriate to the
one-sided junction:
\begin{equation}
{G_d(\omega,V_{\sim})\over G_{d0}^0(V_0)} \simeq F_G^\alpha(\omega)\,
{I_1(\beta V_{\sim})\over \beta V_{\sim}/2}\,,  \qquad
\label{eq:4.77}
\end{equation}
\begin{equation}
{C_d(\omega,V_{\sim})\over C_{d0}^0(V_0)} \simeq F_C^\alpha(\omega)\,
{I_1(\beta V_{\sim})\over \beta V_{\sim}/2}\,.  \qquad
\label{eq:4.78}
\end{equation}
Here we have defined the frequency-dependent factors:
\begin{equation}
F_G^\alpha(\omega) =
r_{p1}^{\alpha}(\omega)\,e^{2\alpha W_n} \equiv
\bigl[ a_{p1}^\alpha(\omega)A_p^\alpha+\alpha L_p \bigr]\,
e^{2\alpha W_n}\!\,,
\label{eq:4.79}
\end{equation}
\begin{equation}
F_C^\alpha(\omega) =
q_{p1}^{\alpha}(\omega)\,e^{2\alpha W_n} \equiv
\frac{1}{a_{p1}^\alpha(\omega)A_p^\alpha}\,
e^{2\alpha W_n}\!\,,  \qquad\quad\,
\label{eq:4.80}
\end{equation}
where\,  $A_p^\alpha= \sqrt{1+ (\alpha L_p)^2}$\,\, and
\[
a_{p1}^\alpha(\omega)A_p^\alpha =\!
\frac{1}{\sqrt2}\,\sqrt{1\!+\!(\alpha L_p)^2\!+
\sqrt{\bigr[1\!+\!(\alpha L_p)^2 \bigr]^2\!+(\omega\tau_p)^2}}\,.
\]
For the uniform PN-junction (with $\alpha =0$) from
Eqs.~(\ref{eq:4.79}) and (\ref{eq:4.80}) it follows that
\begin{equation}
F_G^\alpha(\omega)\,\stackrel{\alpha\to 0\vphantom{j}}{\longrightarrow}\,
{\sqrt{1+\sqrt{1+(\omega\tau_p)^2}}\over\sqrt{2}}\;,
\label{eq:4.81}
\end{equation}
\begin{equation}
F_C^\alpha(\omega)\,\stackrel{\alpha\to 0\vphantom{j}}{\longrightarrow}\,
{\sqrt{2}\over\sqrt{1+\sqrt{1+(\omega\tau_p)^2}}}\;,
\label{eq:4.82}
\end{equation}
so that expressions (\ref{eq:4.77}) and (\ref{eq:4.78}) take a form
identical to formulas (48) and (49) obtained in paper~\cite{BS2007a}.

The dependence on the signal amplitude $V_\sim$ for the diffusion conductance
$G_d(V_\sim)$ and capacitance $C_d(V_\sim)$ appearing in
Eqs.~(\ref{eq:4.77})--(\ref{eq:4.78}) (as well as in the general formulas
(\ref{eq:4.68})--(\ref{eq:4.69})) is expressed by the function
${g_1(V_{\sim})= I_1(\beta V_{\sim})/(\beta V_{\sim}/2)}$.

The frequency dependence of the diffusion conductance $G_d(\omega)$ and
capacitance $C_d(\omega)$ is produced by the functions
$r_{p1}^{\alpha}(\omega)$ and $q_{p1}^{\alpha}(\omega)$. These functions
are defined by Eqs.~(\ref{eq:4.61})--(\ref{eq:4.63}) to appear in
the factors $F_G^\alpha(\omega)$ and $F_C^\alpha(\omega)$ given by
Eqs.~(\ref{eq:4.79}) and (\ref{eq:4.80}). The curves
$r_{p1}^{\alpha}(\omega)$ and $q_{p1}^{\alpha}(\omega)$ are plotted
in Fig.\,1 for different values of the nonuniformity parameter
$\alpha L_p= 0,\,\pm0,5,\,\pm1,\,\pm2,\,\pm3$.

Two curves~1 corresponding to the uniform junction ($\alpha L_p\!=0$) are
fully the same as those shown in Fig.~23 of Chapter~2 in a book by Sze~\cite{SMS1981}.
The cross-sectional nonuniformity ($\alpha L_p\ne 0$) changes the curves
$r_{p1}^{\alpha}(\omega)$ and $q_{p1}^{\alpha}(\omega)$ in different ways.
The function ${q_{p1}^{\alpha}(\omega)<1}$ always and it decreases with growing
$|\alpha L_p|$ regardless of the sign of $\alpha L_p$, as shown by solid
curves $2(2a),\,3(3a),\,4(4a),\,5(5a)$ in Fig.\,1.  By contrast, the~function
$r_{p1}^{\alpha}(\omega)$ increases for $\alpha>0$
(solid curves $2,\,3,\,4,\,5$) and decreases for $\alpha<0$
(dashed curves $2a,\,3a,\,4a,\,5a$) with growing $|\alpha L_p|$,
as compared to curve 1 (for $\alpha=0$).

\begin{figure}[htb]
\includegraphics{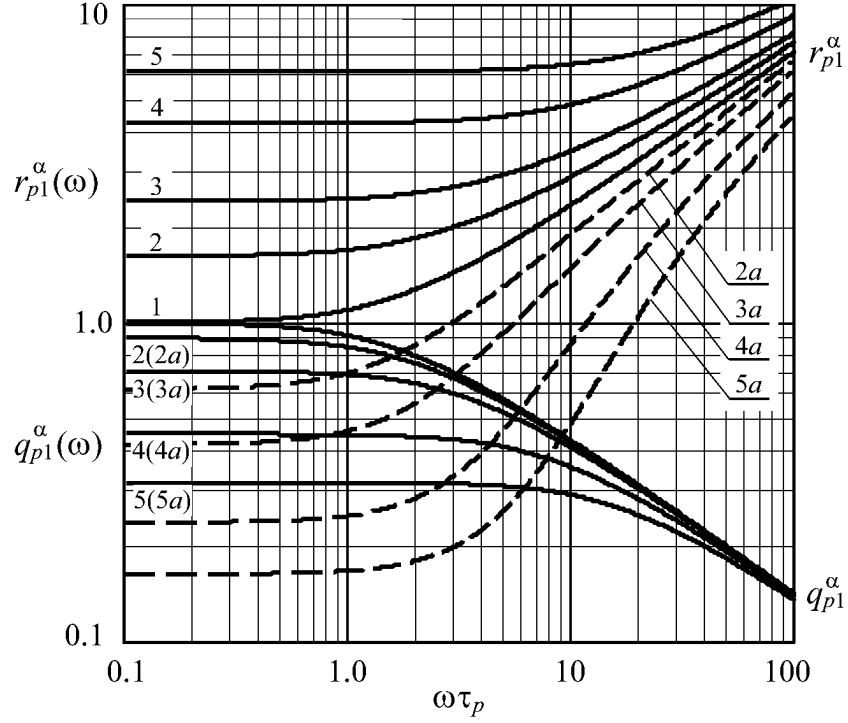}
\caption{Frequency dependencies of the quantities $r_{p1}^\alpha(\omega)$
and $q_{p1}^\alpha(\omega)$ for different values of the nonuniformity
parameter $\alpha L_p= 0$ (solid curves~1), $0.5$ (solid curves~2),
1~(solid curves~3), 2~(solid curves~4), 3~(solid curves~5);
$-0.5$~(dashed curve~$2a$), $-1$~(dashed curve~$3a$),
$-2$~(dashed curve~$4a$), $-3$~(dashed curve~$5a$).}
\label{Fig1}
\end{figure}

At low frequencies such that $\omega\tau_p\!\ll\!1$ and
$a_{p1}^\alpha(\omega)\simeq a_{p1}^\alpha(0)= 1$, the functions
$r_{p1}^{\alpha}(\omega)$ and $q_{p1}^{\alpha}(\omega)$ take
constant values
\[
r_{p1}^{\alpha}(\omega)\simeq r_{p1}^{\alpha}(0)= A_p^\alpha + \alpha L_p
\;\quad\;\mbox{and}\quad\;\;
q_{p1}^{\alpha}(\omega)\simeq q_{p1}^{\alpha}(0)= \frac{1}{A_p^\alpha}\,.
\]
Then, the factors (\ref{eq:4.79}) and (\ref{eq:4.80}) become
frequency-independent and equal to
\begin{eqnarray}
F_G^\alpha(\omega)\simeq F_G^\alpha(0) =
r_{p1}^{\alpha}(0)\,e^{2\alpha W_n} \equiv
(A_p^\alpha+\alpha L_p)\,e^{2\alpha L_p(W_n/L_p)},
\label{eq:4.83}
\end{eqnarray}
\begin{eqnarray}
F_C^\alpha(\omega)\simeq F_C^\alpha(0) =
q_{p1}^{\alpha}(0)\,e^{2\alpha W_n} \equiv
\frac{1}{A_p^\alpha}\;e^{2\alpha L_p(W_n/L_p)}.
\qquad\qquad\quad\;
\label{eq:4.84}
\end{eqnarray}

The dependence of the low-frequency factors (\ref{eq:4.83}) and (\ref{eq:4.84})
on the nonuniformity parameter $\alpha L_p$ is shown in Fig.~2 for three
ratios $W_n/L_p= 0,\,0.05,\,0.1$. Such small values of
$W_n/L_p$ are chosen to ensure the condition ${W_n\ll L_p}$ for neglecting
recombination processes in the depletion layer of width $W_n$~\cite{SMS1981}.
Character of the cross-sectional non\-uniformity (for $\alpha<0$ with
$S_p>S_n$ \,and\, for $\alpha>0$ with $S_p<S_n$) exerts different
influence on $F_G^\alpha(0)$ (or the conductance
$G_d(0,V_\sim)=g_1(V_\sim)F_G^\alpha(0)\,G_{d0}^0$) and on~$F_C^\alpha(0)$
(or the~capacitance $C_d(0,V_\sim)=g_1(V_\sim)F_C^\alpha(0)\,C_{d0}^0$).

The dashed curves in Fig.~2 demonstrate that the low-frequency capacitance
$C_d(0,V_\sim)$ for nonuniform structures (with $\alpha\ne 0$) is
always less than $C_{d0}^0$ for uniform ones (with $\alpha= 0$) for
both signs $\alpha>0$ and $\alpha<0$. But the low-frequency conductance
$G_d(0,V_\sim)$ depicted by solid curves increases for $\alpha>0$
(when $S_n>S_p$) and decreases for $\alpha<0$ (when $S_n<S_p$).

\begin{figure}[htb]
\includegraphics{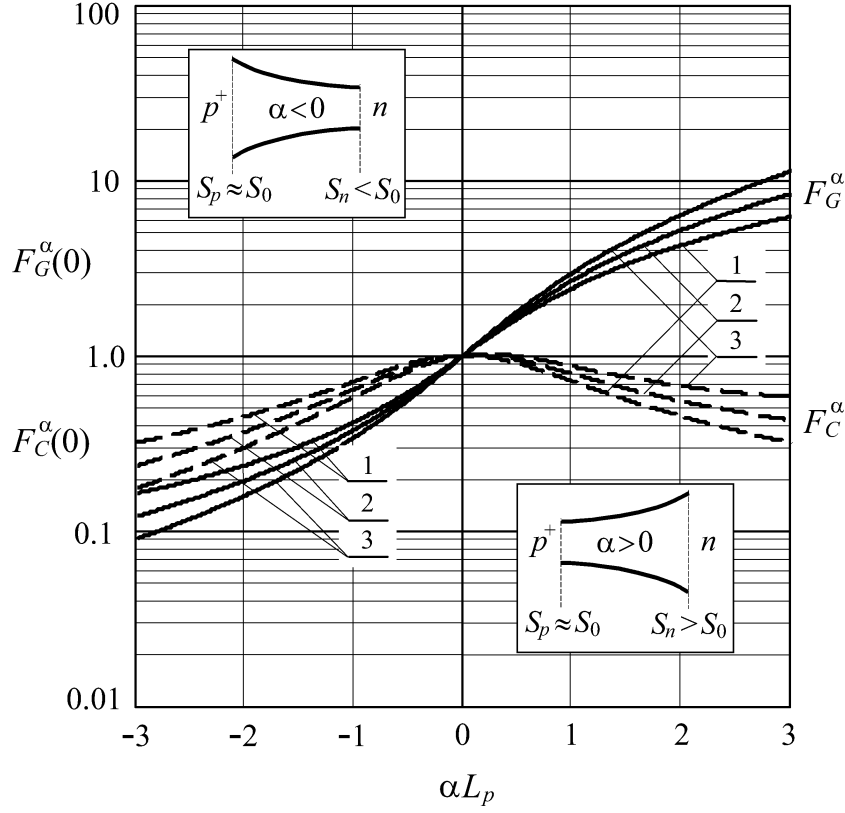}
\caption{Low-frequency values of the factors $F_G^\alpha(0)$
(solid curves) and $F_C^\alpha(0)$ (dashed curves) versus the
nonuniformity parameter $\alpha L_p$ for three values of the ratio
$W_n/L_p= 0$~(curves~1), $0.05$~(curves~2), $0.1$~(curves~3).
The left and right inserts qualitatively show the longitudinal
geometry of a nonuniform structure with $\alpha<0$ and $\alpha>0$.}
\label{Fig2}
\end{figure}

In conclusion, it is pertinent to note that the initial differential
equations (\ref{eq:4.18}) and (\ref{eq:4.19}) have been solved by using
the exponential approximation ${S(z)=S_0\exp(2\alpha z})$ for the cross-sectional
$z$-dependence. In this case, the sought eigenfunctions are of exponential
form $\exp(\pm\Lambda_{pk}^\pm z/L_p)$ and $\exp(\pm\Lambda_{nk}^\pm z/L_n)$
to yield the general solutions (\ref{eq:4.26}) and (\ref{eq:4.27}).
If the power approximation ${S(z)=S_0(\alpha z)^{2m}}$ is more suitable,
Eqs. (\ref{eq:4.18}) and (\ref{eq:4.19}) can be reduced
to the following form (with $u(z)= \Delta\bar p_k(z)$ or
$\Delta\bar n_k(z)$):
\[
\frac{d^2 u(z)}{dz^2}- \frac{2\nu}{z}\,\frac{du(z)}{dz}- c^2 u(z)= 0,
\]
whose solution is (see formula 8.494.9 in Ref.~\cite{GR1980})
\begin{equation}
u(z)= z^{\nu+1/2}Z_{\nu+1/2}(icz),
\label{eq:4.85}
\end{equation}
where $Z_{\nu+1/2}$ is the Bessel function of the first or second kind.
For equations (\ref{eq:4.18}) and (\ref{eq:4.19}) we have $\nu=-m$ and
$c^2=L_k^{-2}$, where $L_k$ is defined by Eq.~(\ref{eq:4.22}). Therefore,
the sought solutions (\ref{eq:4.26}) and (\ref{eq:4.27}) include,
instead of the exponential functions, the new functions (\ref{eq:4.85})
dependent on the complex argument $(iz/L_k)$.

\section{Conclusion}

The paper has demonstrated how to derive the explicit analytic form for
the current--voltage characteristics of the PN-junctions with
nonuniformity in the cross section and doping impurity distribution
by applying the transverse averaging technique (TAT).
Application of the TAT to the three-dimensional transport equations
of semiconductor electronics has converted them into the quasi-1D diffusion
equations (\ref{eq:4.13}) and (\ref{eq:4.14}) to analyze the
minority-carrier transport processes in the nonuniform PN-junctions.

Application of the spectral approach to the quasi-1D diffusion transport
equations for the nonuniform PN-junctions under the action of
arbitrary signal amplitude $V_\sim$ has given rise to changes in both the static
current--voltage characteristic $J_0(V_0)$ and the dynamic characteristics
--- the diffusion conductance $G_d(\omega)$ and capacitance $C_d(\omega)$,
as compared with the conventional theory of uniform junctions~\cite{SMS1981}.
These changes are caused by both factors --- the signal amplitude $V_\sim$
and the nonuniformity of $S(z)$.

The large-signal effects on the static and dynamic characteristics have
proved to be completely identical to those obtained theoretically and
corroborated experimentally for the uniform PN-junctions
in our previous papers~\cite{BS2007a,SB2002}.  As a next step, these 
results should be compared to simulation.

The influence of the cross-sectional nonuniformity on the static
current--voltage characteristic (\ref{eq:4.53}) is exhibited
in terms of the saturation current (\ref{eq:4.54}). The similar influence
on the diffusion conductance $G_d(\omega)$ and capacitance $C_d(\omega)$
is demonstrated by the novel formulas (\ref{eq:4.68})--(\ref{eq:4.69})
and (\ref{eq:4.77})--(\ref{eq:4.78}). The numerical calculations have been
made for the exponential approximation $S(z)=S_0\exp(2\alpha z)$ of the
cross-sectional $z$-dependence.

Until now, large-signal and high-frequency were treated separately to the 
best of our knowledge.  This study may also have an impact in the understanding
of distortion in high frequency circuits.  Further applications, limitations 
of the model, carrier storage effects are the focus of further work.

\section*{Acknowledgment}

One of the authors, AAB, thanks CNPq for
the support during his stay at UFPE.


\bibliographystyle{model1-num-names}
\bibliography{<your-bib-database>}

\begin{thebibliography}{1}
%
\bibitem{SSRF2004} B. Schmith\"{u}sen, A. Schenk, I. Ruiz, and W. Fichtner,
``Simulation of physical semiconductor devices under large and small
signal conditions'', (invited paper), Asia Pacific Microwave Conference -
APMC, New Delhi (2004).
%
\bibitem{YLY1994} A. T. Yang, Y. Liu, and J. T. Yao,
``An efficient nonquasi-static diode model for circuit simulation'',
IEEE Trans. on CAD of Integ. Circ. and Sys., {\bf 13}, 231--239 (1994).
%
\bibitem{RBD1995} R. B. Darling, 
``A full dynamic model for PN-junction diode switching transients'',
IEEE Trans. on Elect. Dev., {bf 42}, 969--976 (1995).
%
\bibitem{RPFAC1993} D. E. Root, M. Pirola, S. Fan, W. J. Anklam, and A. Cognata,
``Measurement-Based Large-Signal Diode Modeling System for Circuit Device
Design'',
IEEE Trans. on Microwave Theory and Techniques, {\bf 41}, 2211--2217 (1993).
%
\bibitem{BS2007a} A. A. Barybin and E. J. P. Santos, `` Transverse averaging
technique for the depletion capacitance of nonuniform PN-junctions'',
Semicond. Sci. Technol. {\bf 22} (2007) 312-319.
%
\bibitem{SB2002} E. J. P. Santos and A. A. Barybin, ``Large-signal
dynamic admittance of $p\!-\!n$-junctions'',  Proc. of the
XVII Int. Symp. on Microelectronics Technology and Devices,
Electrochem. Soc. Proc., PV2002-8 (2002) 237-243.
``Novel Results on the Large-Signal Dynamic Admittance of $p-n$-Junctions'',
cond-mat/0204620.
%
\bibitem{GSBQSB2007} E. Gatard, R. Sommet, P. Bouysse, R. Qu\'{e}r\'{e},
M. Stanislawiak, J.-M. Bureau,
``High Power S Band Limiter Simulation with a Physics-Based Accurate PIN
Diode Model''
Proceedings of the 2nd European Microwave Integrated Circuits Conference,
Munich, 8 to 10 October 2007.
%
\bibitem{SMS1981} S. M. Sze, {\em Physics of Semiconductor Devices\/},
2nd ed. New York: Wiley, 1981.
%
\bibitem{BS2007b} A. A. Barybin and E. J. P. Santos, ``Unified Approach to the
Large-Signal and High-Frequency Theory of PN-Junctions'',
Semicond. Sci. Technol. {\bf 22} (2007) 1225-1231.
%

\bibitem{GR1980} I. S. Gradshteyn and I. M. Ryzhik, {\em Tables of Integrals,
Series, and Products.\/} New York: Academic Press, 1980.
%
\bibitem{Barybin1986} A. A. Barybin, 
{\em Waves in Thin-Film Semiconductor Structures
with Hot Electrons.\/} Moscow: Nauka, 1986 (in Russian).

\end{thebibliography}

\vfil

\pagebreak

\end{document}